\documentclass[prb,twocolumn,showpacs,amssymb]{revtex4}

\usepackage{graphicx}

\begin{document}

\author{Victor Kagalovsky and Demitry Nemirovsky}
\affiliation{Sami Shamoon College of Engineering,
Beer-Sheva, 84100 Israel}

\title{Critical fixed points in class D superconductors}

\begin{abstract}
We study in detail a critical line on the phase diagram of the Cho-Fisher network model 
separating three different phases: metallic and two distinct localized phases with different 
quantized thermal Hall conductances. This system describes non-interacting quasiparticles 
in disordered superconductors that have neither time-reversal nor spin-rotational 
invariance. We find that in addition to a tricritical fixed point $W_T$ on that critical line 
there exist an additional repulsive fixed point $W_N$ (where the vortex disorder concentration 
$W_N<W_T$), which splits RG flow into opposite directions: toward a clean Ising model at $W=0$ and 
toward $W_T$.  
\end{abstract}


\maketitle

The properties of quasiparticles in disordered superconductors belonging to 
new symmetry classes \cite{Zirn}, in particular,  
transitions between metallic, localized, or quantized Hall phases \cite{Sent1, WEPRL, SENT, We}
 , have been intensively studied.
The symmetry class D may be 
realized in superconductors with broken time-reversal and spin-rotation invariances 
(as in d-wave superconductors with spin-orbit 
scattering). The associated changes in quasiparticle dynamics should be examined by 
energy transport, since neither charge density nor spin are conserved. 
In this Brief Report we present a detailed study of the crtitcal line on the phase diagram for a model first introduced by Cho and Fisher (CF) \cite{Cho} (see a detailed description below), which  
has a particularly rich phase diagram in two dimensions \cite{Senthil,We}. 

We employ advanced numerical calculations, proposed in Ref. \onlinecite{We} and described in detail 
in \cite{Merz,Var} to overcome round-off errors in calculations of renormalized localization lengths.  
We also apply an optimization algorithm to determine both critical exponent and critical energy.  
These results allow us to determine the tricritical point where three phases meet and to study dependence of critical exponent for
the insulator-to-insulator transition on the width of the system. Our results suggest the existence of two fixed points.

The original network model \cite{CC} was proposed to describe transitions 
between plateaux in the quantum Hall effect (QHE). In the model flux probabilities move along 
unidirectional links forming closed loops in analogy with semiclassical 
motion of electrons on contours of constant potential. Scattering between links is allowed 
at nodes in order to map tunneling through saddle point potentials. 
Propagation along links is described by diagonal matrices with elements in the form $\exp (i\phi)$. Transfer matrix for one node relates a pair of incoming and outgoing amplitudes 
on the left to a corresponding pair on the right; it has the form
\begin{equation}
{\bf T}=\left( \begin{array}{cc}\sqrt{1+\exp{(-\pi\epsilon)}} & \exp{-(\pi\epsilon/2)}  \\
\exp{(-\pi\epsilon/2)} & \sqrt{1+\exp{(-\pi\epsilon)}}
\end{array}
\right),
\label{first}
\end{equation}
where $\epsilon$ is a dimensionless relative distance between the electron energy and the 
barrier height. It is easy to see that the most "quantum" case (equal
probabilities to scatter to the left and to the right) is at $\epsilon =0$. 
 
Numerical simulations on the network model are performed on a system with fixed width $M$ and periodic boundary conditions in the transverse direction. 
By multiplying transfer matrices for $N$ slices and then diagonalizing the 
resulting total transfer matrix, it is possible to extract the smallest Lyapunov 
exponent $\lambda$ (the eigenvalues of the transfer matrix are $\exp(\lambda N)$). The 
localization length $\xi_M$ is proportional to $1/\lambda$. Renormalized localization lengths
for different system widths and different energies satisfy a one-parameter scaling
\begin{equation}
\frac{\xi_M}{M} =f\left(\frac{M}{\xi (\epsilon )}\right) ,
\label{second}
\end{equation}
which provides dependence of the thermodynamic localization length $\xi$ on energy 
$\epsilon$.

For a D class symmetry 
a Bogoliubov - de Gennes Hamiltonian is written in terms 
of a Hermitian matrix \cite{Zirn}. The corresponding time evolution 
operator is real, and, the generalized phase factors are, therefore, 
O($N$) matrices for a model in which $N$-component fermions propagate on 
links. We study the case $N=1$ with phase factors $\pm 1$. 
There are three models: random bond Ising model \cite{Choth,gruzb}, supporting two different localized phases, 
uncorrelated O(1) model \cite{boucq}, where phases on the links 
are independent random variables and all states are extended \cite{We}, and the model first introduced by Cho and 
Fisher (CF) \cite{Cho} where scattering phases with the value $\pi$ appear 
in correlated pairs. Each model has two parameters: 
the first one is  
a disorder concentration $W$, such that there is a probability $W$  
to have a phase $0$, and a probability ($1-W$) to have a phase $\pi$ on a given link. 
The second parameter is an energy 
$\epsilon$ describing scattering at the nodes. For the CF model, the phase diagram (updated version of which is presented in Fig. 1) in the $\epsilon$-$W$ plane 
has three distinctive phases: metallic, and two insulating phases characterized 
by different Hall conductances. The sensitivity to the disorder is a distinctive 
feature of class D. 

In the CF model the disorder is introduced only at the nodes, allowing for the offdiagonal elements of Eq. (1) to be multiplied by $\pm 1$ (disorder probablity 
$W$ is the probability of that factor to be $-1$). In our previous work \cite{VD} we have studied a CF phase diagram and its critical exponents far from $\epsilon =0$. On the other 
hand, the critical line itself at $\epsilon =0$ is of the particular interest to us. 
One of the possibilities discussed in Ref. \onlinecite{ECM}, where a quasiparticle density of states for the CF model was studied, is the existence of 
the second fixed point (a repulsive one) on the critical line separating two insulating phases at the disorder smaller than the one at the tricitical point, where the critical line splits. It has been suggested 
\cite{FE} that we can address this question using our optimization procedure studying the analogue of RG flow for the critical exponent. Indeed, if one belives the scenario of two 
fixed points: the repulsive one at $W_N$ (the subscript $N$ suggested in Ref. \onlinecite{ECM} 
to underline the similarity to the role of the Nishimori point in the RBIM) and the tricritical 
one at $W_T>W_N$, then the one-parameter scaling for the disorder $W<W_N$ should produce the critical exponents which tend to the value $\nu =1$ of the pure Ising model as one uses only large system widths. In contradistinction, for $W>W_N$ the repulsive fixed point should push the crtitical exponent to flow toward the metallic fixed point.        

We have performed numerical calculations at small fixed values of $W$ for the system widths 
$M=16,32,64,128$ and various energies $\epsilon$. 
In those cases an obvious critical energy is $\epsilon =0$ (the most "quantum" case explained above). We have been determining the critical exponents $\nu$ ($\xi\sim\epsilon^{-\nu}$) to fit all the data onto one curve by applying a special 
optimization program which checks different critical exponents and chooses the optimal one. We have carried out the analysis, first, for all data points for all system widths, second, without $M=16$ data, third, without 
$M=16,32$ data.  
It turns out that, indeed, this procedure shows that the critical exponent flows to the 
predicted pure Ising model value $\nu=1$ as we omit small system widths data (the analogue of RG flow). As a typical example we present here the flow of the critical exponent $\nu$ for the disorder $W=0.04$: $\nu=1.34$ for $M=16,32,64,128$, $\nu=1.11$ for $M=32,64,128$, $\nu=0.97$ 
for $M=64,128$. The largest value of disorder for which this flow persists is $0.1$. 
To demonstrate the flow in a conclusive way at this disorder value, the data for the largest system width $M=256$ was necessary: the critical exponent flows as $1.7\rightarrow 1.6\rightarrow 1.2 \rightarrow 1.1$.     
We, therefore, identify $W_N=0.1$. In a full agreement with the scenario suggested \cite{ECM} for $W>W_N$, the critcal exponent flows in the opposite direction, e.g. for $W=0.12$ the critical exponent is almost constant $\approx 1.7$, independent on the system widths used. 
In order to be sure that we are still on the critical line we allow the oprimization program to look for the critical energy as well. It is very conclusive that up to the disorder value $0.14$ the critical energy is about $10^{-5}$ which supports the single critical line. For $W>0.14$ 
the optimization program immediately produces small but finite values of the critical energy (as shown on Fig. 1). We thus identify the triritical fixed point $W_T=0.14$.

\begin{figure}[htb]
\begin{center}
\includegraphics[scale=0.8]{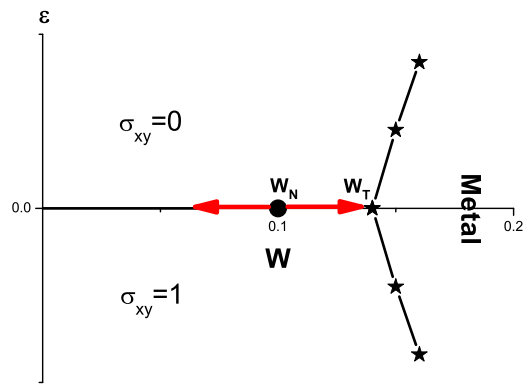}
\end{center}
\caption{Updated phase diagram for a CF model with metallic, insulating and quantized Hall phases.}
\end{figure}

To summarize, we have studied in detail the critical line on the phase diagram of the CF model. 
We have used the RG flow of the critical exponent and determined two fixed points in agreement with one of the scenarios suggested in Ref. \onlinecite{ECM}: the repulsive fixed point at disorder $W_N=0.1$ and the tricritical fixed point $W_T=0.14$.  Obviously, we cannot 
rule out completely that for larger system widths the RG flow can reverse as was found \cite{ECM}.

\begin{acknowledgments}
This research was supported
by the SCE internal research grant. One of us (V.K.) thanks F. Evers, I. Gruzberg and A. Mirlin 
for fruitful discussions.
\end{acknowledgments}

\end{document}